\documentclass{article}
\usepackage{graphicx} 
\usepackage{amsmath}
\usepackage{algorithm}
\usepackage{algpseudocode}
\usepackage{natbib}
\usepackage[utf8]{inputenc}
\usepackage{hyperref}
\usepackage{multirow}
\usepackage{authblk}
\usepackage{xcolor}
\usepackage{setspace}
\title{A Semiparametric Stochastic Volatility Model with Dependent Errors}

\date{}
\author[*]{Yudong Feng}
\author[*]{Ashis Gangopadhyay}
\affil[*]{\textit{Department of Mathematics and Statistics, Boston University, 665 Commonwealth Ave, Boston, MA 02215, USA.}}
\affil[ ]{E-mail: ydfeng@bu.edu, ag@bu.edu}
\begin{document}

\maketitle

\begin{abstract}

This paper proposes a semiparametric stochastic volatility (SV) model that relaxes the restrictive Gaussian assumption in both the return and volatility error terms, allowing them to follow flexible, nonparametric distributions with potential dependence. By integrating this framework into a Bayesian Markov Chain Monte Carlo (MCMC) approach, the model effectively captures the heavy tails, skewness, and other complex features often observed in financial return data. Simulation studies under correlated Gaussian and Student’s t error settings demonstrate that the proposed method achieves lower bias and variance when estimating model parameters and volatility compared to traditional Gaussian-based and popular Bayesian implementations. We conduct an empirical application to the S$\&$P 500 index, which further underscores the model’s practical advantages: it provides volatility estimates that respond more accurately to large fluctuations, reflecting real-world market behavior. These findings suggest that the introduced semiparametric SV framework offers a more robust and adaptable tool for financial econometrics, particularly in scenarios characterized by non-Gaussian and dependent return dynamics.

\smallskip

\noindent \textbf{Keywords:} Stochastic Volatility Model, Bayesian Inference, Nonparametric Method, Markov Chain Monte Carlo, Volatility Modeling, Financial time series.

\end{abstract}

\smallskip

\section{Introduction}

In financial time series analysis, the volatility of asset returns constitutes a fundamental area of research. The importance of volatility modeling is underscored by its extensive applications in key financial domains, including asset pricing, portfolio optimization, and risk management, including the Markowitz portfolio allocation framework by \cite{Markowitz} and the Black-Scholes option pricing model by \cite{BS}. Due to its critical significance, volatility estimation has been a focus point of financial time series research for over three decades. 

Subsequent investigations employing similar model frameworks have been conducted by numerous researchers, including the works by \cite{Hull_White}, \cite{Chesney_Scott}, \cite{Taylor}, \cite{Jacquier}, \cite{Jacquier2002}, and \cite{Shephard}, focusing on both parametric and semiparametric estimation techniques for stochastic volatility models. \cite{Harvey1994} introduced an asymmetric stochastic volatility model that incorporates the leverage effect through a quasi-maximum likelihood estimation procedure within a state-space framework, thereby enabling the analysis of financial time series exhibiting asymmetric volatility patterns. \cite{Carter_Kohn} advanced a Gibbs sampling approach for state-space models, demonstrating particular effectiveness in settings where errors follow mixture distributions and coefficients evolve over time. \cite{Kim1998} proposed a Bayesian treatment of stochastic volatility using MCMC methods, emphasizing the advantages of SV models over ARCH models in capturing the volatility dynamics of financial time series and providing methodologies for filtering, diagnostics, and model selection. Furthermore, \cite{Omori2007} extended Bayesian analyses of SV models to account for leverage effects and introduced a robust MCMC-based method that utilizes a mixture approximation for the joint innovation distributions.

\cite{Taylor}'s Stochastic Volatility (SV) Model has been widely recognized in financial econometrics for its capacity to model time-varying volatility. The Taylor style model of returns $y_1, y_2,..., y_t$ is
\begin{equation}
    \begin{aligned}
        y_t & =\sqrt{h_t}u_t,\  t = 1,...,N \\
        \ln h_t & =\alpha+\delta\ln h_{t-1}+\sigma_\nu\nu_t,\ t = 2,...,N
    \end{aligned}
    \label{eq: model}
\end{equation}
where $u_t,\nu_t\sim N(0,1)$ for $t = 1,...,T$ and $corr(u_t,\nu_{t'}) = 0$ for any $t$ and $t'$.

Nonetheless, as noted by \cite{Durham2006}, the model falls short in representing the heavy-tailed characteristics frequently observed in the conditional distribution of returns, primarily due to its dependence on the Gaussian error assumption. This limitation arises from the reliance on a Gaussian assumption for the error terms. In financial econometrics, a central objective is to analyze the distribution of returns in financial markets, as it provides critical insights into the mechanisms driving financial dynamics. In his seminal study of cotton price fluctuations, \cite{Mandelbrot1963} identified significantly heavier tails than those implied by the normal distribution, thereby motivating the investigation of stable distributions as a more appropriate alternative to the traditional Gaussian-based models.

To tackle this issue, researchers like \cite{Jacquier2004} have proposed alternative parametric heavy-tailed distributions, namely the Student’s t distribution with low degrees of freedom or the Generalized Error Distribution (GED), as potential remedies. These distributions offer greater flexibility than the Gaussian assumption by accommodating the heavy-tailed characteristics commonly observed in empirical financial data. Consequently, they promise to provide more accurate risk estimates and volatility forecasts, particularly in contexts where extreme events or tail dependence are of principal concern.

However, as emphasized in the work of \cite{efficiency_semi_para} and \cite{One-step}, the parametric approaches can still be inadequate when the true error distribution is not clearly specified. Despite their capacity to capture heavier tails than the normal distribution, such assumptions may oversimplify the inherent complexity of the data and fail to represent crucial structural features in financial returns. This shortcoming becomes especially apparent in real-world scenarios where data exhibit idiosyncratic behaviors, including asymmetry, time-varying dynamics, or other departures from standard parametric distributions. Consequently, the limitations of these models have stimulated continued research into alternative, more flexible methods that aim to offer a more comprehensive framework for modeling heavy-tailed phenomena in finance.

A feasible strategy to address this limitation is to employ a nonparametric distribution for the return error term, thereby affording greater flexibility in capturing the empirical features of financial data and effectively allowing the data to “guide the distributional assumption”. Similar ideas have been investigated by \cite{One-step} and \cite{efficiency_semi_para} in the context of GARCH models, where the authors introduced a nonparametric estimation procedure for the error distribution. This advancement led to the construction of a pseudo-likelihood function for the model, with the resultant volatility estimates derived from the GARCH parameters obtained through this pseudo-likelihood approach.

This paper extends the methodology introduced by \cite{semiSVM} by adopting fully nonparametric distributions for error terms, $u_t$ and $\nu_t$. This approach further relaxes the underlying assumptions regarding the distribution of these error components, thereby offering greater flexibility in capturing potential non-Gaussian features, such as skewness and heavy tails, often observed in financial time series. By removing the need to specify a particular parametric form, the proposed framework can more accurately reflect the inherent complexities of the data, ultimately improving the robustness of volatility modeling.

\section{A semiparametric approach to SVM}

In conventional parametric SV models, the error terms $u_t$ and $\nu_t$ are typically assumed to follow independent normal distributions. However, the Gaussian assumption has faced increasing scrutiny and challenge in recent literature. For instance, \cite{Omori2007} has advocated for using a Student-t distribution, while \cite{Barndorff-Nielsen} proposed the Normal-Inverse Gaussian distribution as an alternative. Furthermore, \cite{Mahieu_Schotman} employed a mixture of normal distributions to better capture the underlying volatility dynamics, and \cite{Abanto-Valle2011} applied a scale mixture of normals with varying mixing parameters. Notably, \cite{One-step} and \cite{Data-Dependent} highlighted the limitations of parametric assumptions in GARCH models, emphasizing that such models often fail to accurately reflect the distributions observed in financial data. Collectively, these studies underscore a growing shift toward adopting more diverse and representative error distributions for modeling volatility in financial time series. Besides the parametric assumption of the error term, the assumption of independent structure of the errors $u_t$ and $\nu_t$ is also questioned recently. \cite{Jacquier2004} suggested using $\rho = corr(u_t, \nu_t)$ to capture the correlation relation. \cite{Chan_Kohn_Kirby} developed a Bayesian approach for parsimoniously estimating the correlation of errors. \cite{Mukhoti_Ranjan} introduced a Bayesian framework for efficiently estimating the correlation structure of errors in multivariate SV model using a sparsity-inducing prior and MCMC methods. 

\cite{semiSVM} introduces innovative semiparametric approaches to stochastic volatility (SV) modeling, addressing limitations in traditional parametric models, particularly their reliance on normally distributed errors. The study provides compelling evidence of the advantages of semiparametric methods in capturing heavy-tailed distributions and improving volatility estimation accuracy, making significant contributions to financial econometrics.

 The current investigation begins by identifying the shortcomings of traditional SV models, such as their inability to account for extreme returns observed in financial data. It then proposes two semiparametric SV models, which leverage kernel density estimation to relax the Gaussian error assumption. This methodological innovation allows for greater flexibility in modeling real-world financial time series, especially under heavy-tailed error distributions like the Student-t and Generalized Error Distributions (GED).

Using Bayesian inference techniques and Markov Chain Monte Carlo (MCMC) sampling, \cite{semiSVM} systematically estimates model parameters and compares the performance of the semiparametric and traditional Gaussian models. The simulations demonstrate that the semiparametric models exhibit lower bias and variance, outperforming their parametric counterparts in both Gaussian and heavy-tailed error scenarios. The empirical application to S$\&$P 500 data further reinforces the practical relevance of these models, showing superior adaptability to market dynamics.

The use of kernel density estimation is particularly noteworthy in the work by \cite{semiSVM}. Utilizing the residuals to develop a semiparametric model for the error distributions, the paper ensures that the error terms better reflect the empirical distribution of financial returns. This refinement is crucial for asset pricing, portfolio optimization, and risk management applications, where accurate volatility estimation is paramount.

While it provides robust theoretical and empirical validation of its models, certain areas could benefit from further exploration. For instance, in practical applications, the error terms in Equation \ref{eq: model} may exhibit statistical correlation, potentially violating the underlying assumptions of the proposed framework.

\section{A Bayesian model framework}

The Taylor-style stochastic volatility model by \cite{Taylor} is 
\begin{equation}
    y_t=\sqrt{h_t}u_t,\  t = 1,...,N 
    \label{eq: return}
\end{equation}
\begin{equation}
    \ln h_t=\alpha+\delta\ln h_{t-1}+\sigma_\nu\nu_t,\ t = 2,...,N
    \label{eq: vol_eq}
\end{equation}

For the given $t = 1,...,N$, the return of an underlying financial asset is $y_t$ and the latent log volatility $h_t$ follows the first-order auto-regressive process with $\delta$, $\alpha$ and $\sigma_\nu$ parameters, where $\alpha$ is the intercept parameter, $\delta$ is the volatility persistence and $\sigma_\nu$ is the standard deviation of the shock to the logarithm of $h_t$. 

This specific formulation of the model has been utilized in various studies, including those by \cite{Taylor}, \cite{Hull_White}, \cite{Chesney_Scott}, \cite{Shephard}, \cite{Ghysels_Harvey_Renaul}, \cite{Jacquier}, and \cite{Kim1998}. These works provide an in-depth examination of the fundamental econometric properties of the model and the methodologies employed for estimating Stochastic Volatility models.

If we choose the prior of $\delta$, $\alpha$, $\sigma_\nu$ as $p(\delta)\sim N(\delta_0,\sigma_\delta^2)$, $p(\alpha)\sim N(\alpha_0,\sigma_\alpha^2)$ and $p(\sigma_\nu^2)\sim Inv \textbf{-} Gamma(\frac{\nu_0}{2},\frac{s_0}{2})$, by the work of \cite{Jacquier}, the joint likelihood of the parametric SVM \ref{eq: model} by \cite{Taylor} is 
\begin{equation}
\label{eq:likelihood}
    \begin{aligned}
        p(y,h,\delta,\alpha,\sigma_\nu^2)\propto & \frac{1}{\sigma_\nu^{\nu_0}\sigma_\delta\sigma_\alpha\sigma_\nu^2}\exp(-\frac{(\delta-\delta_0)^2}{2\sigma_\delta^2}-\frac{(\alpha-\alpha_0)^2}{2\sigma_\alpha^2}-\frac{s_0^2}{2\sigma_\nu^2}) \\
        & \times\prod_{t=2}^{N}\frac{1}{h_t^{\frac{3}{2}}\sigma_\nu}\exp(-\frac{y_t^2}{2h_t}-\frac{(\ln h_t-\delta\ln h_{t-1}-\alpha)^2}{2\sigma^2_\nu})
    \end{aligned}
\end{equation}

In equation \ref{eq:likelihood}, we choose the prior of $\delta$, $\alpha$, $\sigma_\nu$ as $p(\delta)\sim N(\delta_0,\sigma_\delta^2)$, $p(\alpha)\sim N(\alpha_0,\sigma_\alpha^2)$ and $p(\sigma_\nu^2)\sim Inv \textbf{-} Gamma(\frac{\nu_0}{2},\frac{s_0}{2})$.

From this joint distribution, we can derive the posterior distributions of $\delta$, $\alpha$, $\sigma_\nu$:

\begin{equation}
    p(\delta|h,\alpha,\sigma_\nu^2)\sim N\left(\frac{\sigma_\nu^2\delta_0+\sigma_\delta^2(s_3-\alpha s_1+\alpha\ln h_N)}{\sigma_\nu^2+\sigma_\delta^2(s_2-(\ln h_N)^2)},\frac{\sigma_\nu^2\sigma_\delta^2}{\sigma_\nu^2+\sigma_\delta^2(s_2-(\ln h_N)^2)}\right)
\end{equation}
\begin{equation}
    p(\alpha|h,\sigma_\nu^2,\delta)\sim N\left(\frac{\sigma_\nu^2\alpha_0+\sigma_\alpha^2((1-\delta)s_1-\ln h_1+\delta\ln h_N)}{\sigma_\nu^2+N\sigma_\alpha^2-\sigma_\alpha^2},\frac{\sigma_\nu^2\sigma_\alpha^2}{\sigma_\nu^2+N\sigma_\alpha^2-\sigma_\alpha^2}\right)
\end{equation}
\begin{equation}
    p(\sigma_\nu^2|h,\alpha,\delta)\sim IG\left(\frac{\nu_0+N-1}{2},\frac{s}{2}\right)
\end{equation}
where
\begin{equation}
    s=s_0+(N-1)\alpha^2+(1+\delta^2)s_2-\delta^2(\ln h_N)^2-(\ln h_1)^2-2\alpha(s_1\delta s_1-\ln h_1+\delta\ln h_N)-2\delta s_3
\end{equation}
\begin{equation}
    s_1=\sum_{t=1}^N\ln h_t, 
    s_2=\sum_{t=1}^N(\ln h_t)^2,
    s_3=\sum_{t=2}^N \ln h_t\ln h_{t-1}
\end{equation}

Then, we derive that the conditional posterior of $h_t$ is
\begin{equation}
    \label{eq:5}
    p(h_t|h_{t+1},h_{t-1},\delta,\alpha,\sigma_\nu^2)\propto\frac{1}{\sqrt{h_t}}\exp\left(-\frac{y_t^2}{2h_t}\right)\frac{1}{h_t}\exp\left(-\frac{(\ln h_t-\mu_t)^2}{2\sigma^2}\right)
\end{equation}

where
\begin{equation}
    \mu_t=\frac{\delta\ln h_{t+1}+\delta\ln h_{t-1}+(1-\delta)\alpha}{1+\delta^2}, 
    \sigma^2=\frac{\sigma_\nu^2}{1+\delta^2}
\end{equation}

Inspired by the articles by \cite{One-step}, \cite{Data-Dependent}, and \cite{semiSVM}, we assume the error term in the return equation \ref{eq: return}, $u_t$ follows a nonparametric distribution with density $f$, and the error term in the volatility equation \ref{eq: vol_eq}, $\nu_t$ follows another nonparametric distribution with density $g$. Given the assumption that correlation between the two error terms, $\rho(u_t, \nu_t) \neq 0$, we choose to use a bivariate density, namely, $k$ to represent the dependence via the joint density of $u_t$ and $\nu_t$, i.e., $p(u_t,\nu_t) \sim k$. Similar to the work by \cite{semiSVM}, we leverage the bivariate kernel density estimation technique to approximate the nonparametric $k$. 

The work of \cite{Jacquier} and \cite{Jacquier2004} introduced a Bayesian framework for parameter estimation and volatility sampling, assuming normally distributed error terms. This approach facilitates the generation of volatility samples, represented as $\hat{h_0}, \hat{h_1}, ..., \hat{h_N}$, as well as the posterior distributions of the model parameters: $\delta$, $\alpha$, and $\sigma_\nu$. Consequently, the Bayesian estimates of the parameters, denoted as $\hat{\delta}$, $\hat{\alpha}$, and $\hat{\sigma_\nu}$, are computed as the means of their respective posterior distributions. Let $y_t$ represent the asset's return, we define $\hat{u_t}$ as $\hat{u_t} = \frac{y_t}{\sqrt{h_t}}$, $t = 0,1,...,N$, and $\hat{w_t}$ as $\hat{w_t} = \frac{\ln h_t - \hat{\mu_t}}{\hat{\sigma_\nu}}$, $t=0,1,...,N$, where $\hat{\mu_t} = \frac{\hat{\delta}\ln \hat{h_{t+1}}+\hat{\delta}\ln \hat{h_{t-1}}+(1-\hat{\delta})\hat{\alpha}}{1+\hat{\delta}^2}$, as the residuals of the return and volatility equations, based on the sampled volatilities and the parameter estimates from Gaussian SVM. We replace the two Gaussian components of the posterior in Equation \ref{eq:5} with the kernel density estimate of $h$, namely $\hat{k}$, that takes the standardized residuals $\hat{u_t}$ and $\hat{w_t}$, i.e., $\frac{\hat{u_t} - mean(\hat{u_t})}{sd(\hat{u_t})}$ and $\frac{\hat{w_t} - mean(\hat{w_t})}{sd(\hat{w_t})}$ as input. Thus, if we assume the error terms $u_t$ in Equation \ref{eq: return} and $\nu_t$ in Equation \ref{eq: vol_eq} are nonparametrically distributed, dependent, and can be evaluated by kernel density estimation, then the approximation of the posterior of volatility $h_t$ is 
\begin{equation}
    p(h_t|h_{t+1},h_{t-1},\delta,\alpha,\sigma_\nu^2)\propto \frac{1}{h_t^{3/2}}\hat{k}\left(\frac{y_t}{\sqrt{h_t}}, \frac{\ln h_t-\mu_t}{\sigma_\nu}\right)
    \label{eq:nonpara_vol_likelihood}
\end{equation}
where $\hat{h}(x,y) =  \frac{1}{n b_x b_y} \sum_{i=1}^n N\left(\frac{x - x_i}{b_x}, \frac{y - y_i}{b_y}\right)$, here $b_x$ and $b_y$ are bandwidth of variable $x$ and $y$ respectively, and here $N$ is the kernel function. We name the stochastic volatility model under this assumption as NSVM-3, as a further step of NSVM-1 and NSVM-2 in the work by \cite{semiSVM}. For implementation, we utilize \texttt{kde2d} function from package \textsc{MASS} in R by \cite{MASS} to obtain the kernel density estimation of Equation \ref{eq:nonpara_vol_likelihood} and the bandwidth $b$ is chosen optimally using a standard bandwidth selection method.

\section{Algorithm for Bayesian SVM}

As is commonly observed, the nonparametric assumption for residuals results in the posterior distribution of $h_t$ lacking a closed-form expression. Consequently, numerical methods are required to sample from this posterior distribution for the estimation of both volatility and model parameters. In this study, we employ the Metropolis-Hastings (MH) algorithm mentioned in the work by \cite{MH}, \cite{MH1}, \cite{MH2}, and \cite{MH3} to address the challenges associated with intractable posterior distributions. This approach also facilitates the design of tailored proposal distributions specific to the model, thereby improving the efficiency of the sampling process.

Algorithm \ref{alg:sketch_main} presents a sketch of the primary algorithm. 

\begin{algorithm}[H]
    \caption{Sketch main algorithm} 
    \begin{algorithmic}
        \Require $Input: (y,v_0,s_0,\delta_0,\sigma_\delta^2,\alpha_0,\sigma_\alpha^2,T,b)$
        \State Initialize parameters $\delta$, $\alpha$, $\sigma_\nu$ and $h$
        \For{i = 1,...,T+b}
            \For{t = 1,...,N-1}
                \State Draw $h_t$ from its posterior distribution
            \EndFor
            \State Initialize $h_0$ be the value such that $ln\ h_1 = \alpha + \delta ln\ h_{0} + \sigma_\nu \nu_1$
            \State Let $ln\ h_N = \alpha + \delta ln\ h_{N-1} + \sigma_\nu \nu_N$
            \State Draw $\delta$ from its posterior distribution
            \State Draw $\alpha$ from its posterior distribution
            \State Draw $\sigma_\nu$ from its posterior distribution
        \EndFor
    \end{algorithmic}
    \label{alg:sketch_main}
\end{algorithm}

\subsection{Semiparametric SVM algorithm}

Semiparametric Bayesian models provide significant flexibility by permitting model complexity to increase with the size of the data, thereby adapting to the underlying structure without requiring predefined assumptions regarding the number of parameters. This adaptability is particularly beneficial in situations where the true structure of the data is unknown, as it allows the model to capture intricate and complex patterns that might be missed by parametric models.

The sampling method proposed by \cite{Jacquier} offers an efficient technique to the updating step of the Metropolis-Hastings (MH) algorithm by \cite{MH} and \cite{MH2}. This improvement is based on an informed approximation of a target distribution using an inverse gamma distribution. The choice of the inverse gamma distribution is motivated by its ability to align the first and second moments of the proposed distribution with those of the log-normal component in the posterior distribution.

The proposal distribution $q(h_t)$ is defined by the equation:
\begin{equation}
    q(h_t)=\frac{\lambda^\phi}{\Gamma(\phi)}h^{-(\phi+1)}e^{-\frac{\lambda}{h_t}}
\end{equation}

where
$
\lambda=\frac{1}{2}+\frac{1-2e^{\sigma^2}}{1-e^{\sigma^2}}
$ and 
$
\phi=\frac{y_t^2}{2}+(\lambda-1)e^{\mu_t+\frac{\sigma^2}{2}}
$ are the parameters of the inverse gamma distribution.

The sketch algorithm for sampling the volatility is shown in Algorithm \ref{alg:sketch_sampler_h}.

\begin{algorithm}[H]
    \caption{Sketch algorithm of sampler for $h$}  
    \begin{algorithmic}
        \Require $Input: (y,\alpha,\delta,\sigma_\nu^2,ln\_h_{i-1},h_i,ln\_h_{i+1})$
        \State Calculate $\mu$, $\sigma^2$, $\lambda$, $\phi$ by using input values
        \While{not accepted}
            \State Draw the proposed $h_{new}\sim IG\left( \lambda,\phi \right)$
            \State Accept $h_{new}$ with probability $Min\left(1,\frac{p(h_{new})}{c q(h_{new})} \right)$, where $q$ is the density of inverse gamma distribution.
        \EndWhile
        \If{$p(h_{new})<c q(h_{new})$}
            \State \Return $h_{new}$
        \EndIf
        \State Accept $h_{new}$ with probability $Min\left(1,\frac{p(h_{new})/q(h_{new})}{p(h_{old})/q(h_{old})} \right)$, where $h_{old}$ is previous sample of $h_t$
        \If{accepted}
            \State \Return $h_{new}$
        \Else 
            \State \Return $h_{old}$
        \EndIf
    \end{algorithmic}
    \label{alg:sketch_sampler_h}
\end{algorithm}

Here, the constant $c$ is defined as:
\begin{equation}
    \label{eq:c_star}
    c=c_\star\left(\frac{p(h_m)}{q(h_m)}\right)
\end{equation}
 
where $h_m$ is the mode of $q$ and $c_\star$ here is a tuning parameter. This constant serves as a scaling factor, ensuring that the proposed distribution $q(h_t)$ closely mirrors the target distribution $p(h)$ in terms of its mode. In our implementation in the simulation, we choose $c_\star = 1.2$.

If we assume $u_t$ and $\nu_t$ are nonparametrically distributed and do not necessarily need to be independent, we need numerical methods like MCMC to sample from their nonparametric posterior distributions as shown in the paper by \cite{semiSVM}. The approach can be extended to the present scenario where the errors are dependent and their joint distribution characterizes the dependence. Here we use parameter $\delta$ as an example; similar sampling procedures are deployed on other parameters.

The sketch of the sampler algorithm for $\delta$ in the semiparametric setup is shown in Algorithm \ref{alg:sketch_sampler_delta}.

\begin{algorithm}[H]
    \caption{Sketch algorithm of sampler for $\delta$}
    \begin{algorithmic}
        \Require $Input: (\delta_0,\sigma_\delta^2,\alpha,\sigma_\nu^2,h,\delta_{left})$
        \State Let proposal distribution $q$ be the same as its prior distribution
        \State Let $p$ be the posterior distribution in a parametric setting
        \While{not accepted}
            \State Draw proposed $\delta_{new}$ from proposal distribution
            \State Accept $\delta_{new}$ with probability $Min\left(1,\frac{p(\delta)}{cq(\delta)}\right)$
        \EndWhile
        \If{$p(\delta)<cq(\delta)$}
            \State \Return $\delta_{new}$
        \EndIf
        \State Accept $\delta_{new}$ with probability $Min\left(1, \frac{p(\delta_{new})/q(\delta_{new})}{p(\delta_{old})/q(\delta_{old})} \right)$, where $\delta_{old}$ is the previous sample of $\delta$
        \If{accepted}
            \State \Return $\delta_{new}$
        \Else
            \State \Return $\delta_{old}$
        \EndIf
    \end{algorithmic}
    \label{alg:sketch_sampler_delta}
\end{algorithm}

\section{A comparison of model performances}

This section presents the results of applying the proposed model to simulated data. In real-world applications, the true values of volatility and model parameters are typically unobservable, making it difficult to evaluate performance differences between parametric and semiparametric models. Therefore, we utilize simulations that can provide a known ground truth for parameter values and realized volatility, allowing us to obtain a more precise and detailed comparison of model performance. To access our performance, we compare it with the method \textsc{stochvol} proposed by \cite{stochvol}, a \textsc{R} package that offers a comprehensive Bayesian framework for modeling heteroskedasticity in time series data through stochastic volatility models. It facilitates inference by employing advanced Markov Chain Monte Carlo algorithms to generate samples from the joint posterior distribution of model parameters and latent volatility states.

\subsection{Gaussian model with dependent errors}

We utilize the data-generating process that accounts for the dependence between the error terms in the following manner.

Initialize $h_0\sim \mathrm{N}\left(\mu=-10,\frac{\sigma}{\sqrt{1-\delta^2}}=0.87\right)$, $y_0 \sim \sqrt{h_0} * \mathrm{N}\left(0,1\right)$. For $i = 1,2,...,500$, $y_i =\sqrt{h_i}u_i$, $ln\left(h_{i}\right) = \alpha + \delta ln\left(h_{i-1}\right) + \sigma_\nu \nu_i$. Here $u_i$ and $\nu_i$ are bivariate Gaussian distribution with $\mathrm{mean}(u_i)=\mathrm{mean}(\nu_i)=0$, $\mathrm{var}(u_i)=\mathrm{var}(\nu_i)=1$ and $\rho(u_i,\nu_i) = -0.5$. For the Gaussian SVM and NSVM-3 models, we run each MCMC for 10,000 iterations with the first 5,000 iterations designated as burn-in. Only the samples obtained after the 5,000th iteration were retained for subsequent analysis. For the usage of \textsc{stochvol}, we use the same prior distribution for priors and the same data set that has been generated.

To evaluate the precision and accuracy of parameter estimation, the process was repeated 50 times. In each iteration, we generate a new dataset, $y_1, ..., y_{500}$ using the same parameter values, and the corresponding results: mean, median, and mode of the posterior distributions, $p(\alpha|h,\delta,\sigma_\nu^2)$, $p(\sigma_\nu^2|h,\alpha,\delta)$, and $p(\delta|h,\alpha,\sigma_\nu^2)$ were recorded.

\begin{figure}[H]
    \centering
    \includegraphics[width=\linewidth]{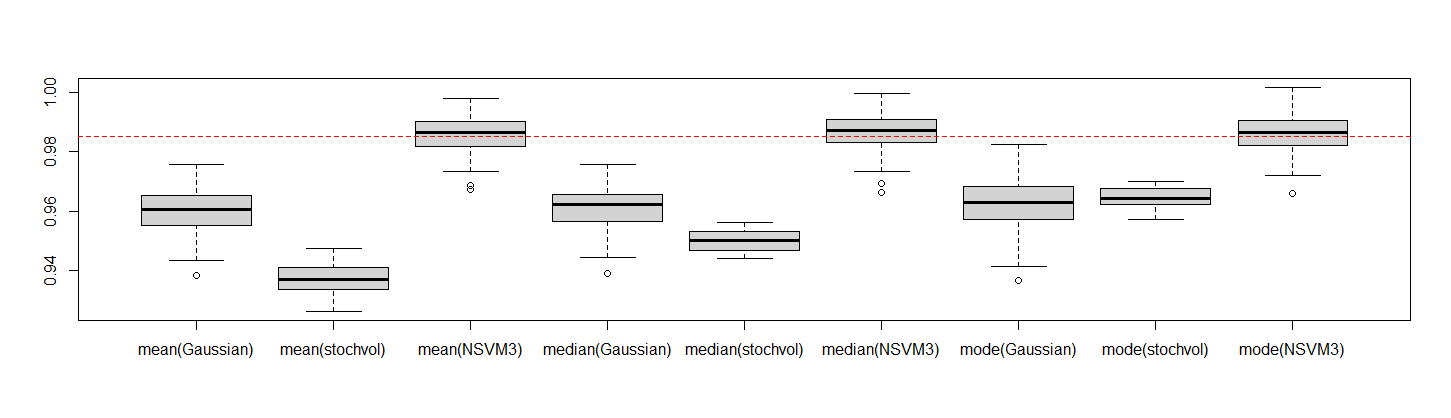}
    \caption{Delta estimation (Dependent Gaussian error simulation)}
    \label{fig:100run_delta_gaussian}
\end{figure}

\begin{figure}[H]
    \centering
    \includegraphics[width=\linewidth]{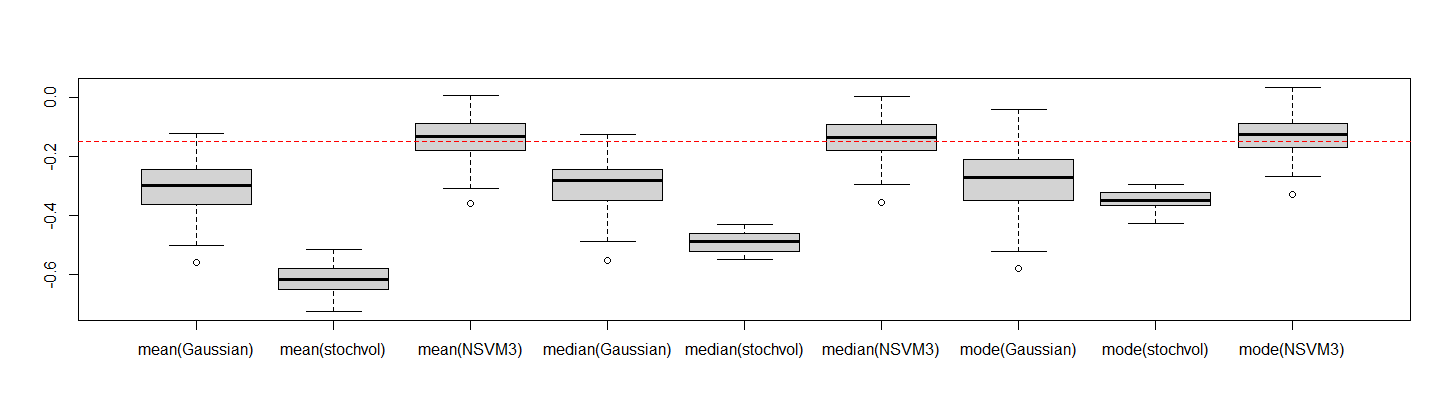}
    \caption{Alpha estimation (Dependent Gaussian error simulation)}
    \label{fig:100run_alpha_gaussian}
\end{figure}

\begin{figure}[H]
    \centering
    \includegraphics[width=\linewidth]{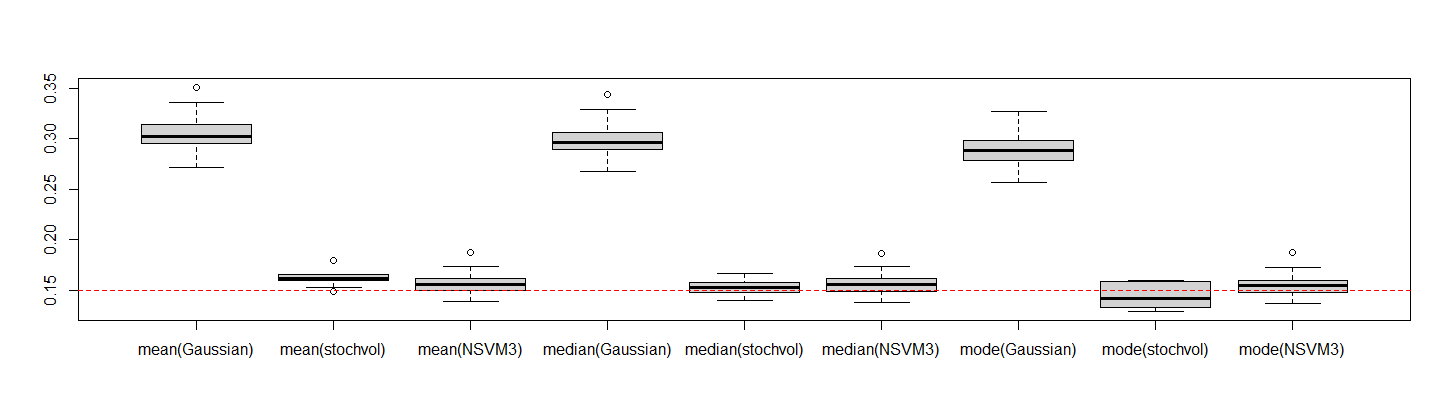}
    \caption{Sigma estimation (Dependent Gaussian error simulation)}
    \label{fig:100run_sigma_gaussian}
\end{figure}

As illustrated in the box plots, the results obtained from the NSVM-3 model exhibit a smaller deviation from the red dashed line, which stands for the true value of the corresponding parameter, compared to those of the Gaussian model. This observation suggests that the semiparametric model demonstrates lower bias. Furthermore, the narrower range of the box plots for NSVM-3 indicates substantially lower variance in the estimates. To quantitatively assess performance, we define the square root mean square error (srMSE) of them can be defined as $\sqrt{\frac{1}{100}\left(\Sigma \bar{\delta_i} - \delta \right)^2}$, $\sqrt{\frac{1}{100}\left( \Sigma \bar{\alpha_i} - \alpha \right)^2}$ and $\sqrt{\frac{1}{100}\left( \Sigma \bar{\sigma_{\nu i}} - \sigma_\nu \right)^2}$. Table \ref{tab:para_est_compare_Gaussian} shows the srMSE of mean, median, and mode of the posterior of parameters $\delta$, $\alpha$, and $\sigma_\nu$ from 100 runs.

\begin{table}[H]
    \centering
    \resizebox{\textwidth}{!}{
    \begin{tabular}{|c|c|c|c|c|c|c|c|c|c|c|}
        \hline
        srMSE  & $\delta$ Mean & $\delta$ Median & $\delta$ Mode & $\alpha$ Mean & $\alpha$ Median & $\alpha$ Mode & $\sigma_\nu$ Mean & $\sigma_\nu$ Median & $\sigma_\nu$ Mode \\
        \hline
        Gaussian &  0.02633411  & 0.02550775 & 0.02515998 & 0.1817484 & 0.1729169 & 0.1762524 & 0.1557682 & 0.1493876 & 0.1389039 \\
        stochvol  &  0.04799186  & 0.03528350 & 0.02098708 & 0.4669827 & 0.3420457 & 0.2038468 & 0.0142296 & \textbf{0.0080419} &0.0139214 \\
        NSVM-3  &  \textbf{0.0058828}  & \textbf{0.0064611} & \textbf{0.00695156} & \textbf{0.0696419} & \textbf{0.0679495} & \textbf{0.0711801} & \textbf{0.01098193} & 0.01040557 & \textbf{0.0096201} \\
        \hline
    \end{tabular}}
    \caption{Parameter estimation comparison, dependent Gaussian error simulation (lower is better)}
    \label{tab:para_est_compare_Gaussian}
\end{table}

As the table \ref{tab:para_est_compare_Gaussian} shows, the NSVM-3 model achieves the lowest square root mean squared error (srMSE) across all metrics, with the exception of the median of the $\sigma_\nu$ posterior distributions. These results indicate that the semiparametric approach offers a clear advantage in reducing bias in parameter estimation.

\subsection{Dependent Student's t distributed errors}

In our simulation, we set $u_t$ and $\nu_t$ to follow a bivariate Student-t distribution with 10 degrees of freedom. In implementation, we utilize \texttt{rmvt} function to generate $u_t$ and $\nu_t$ from \textsc{mvnfast} package in \textsc{R}.

\begin{figure}[H]
    \centering
    \includegraphics[width=\linewidth]{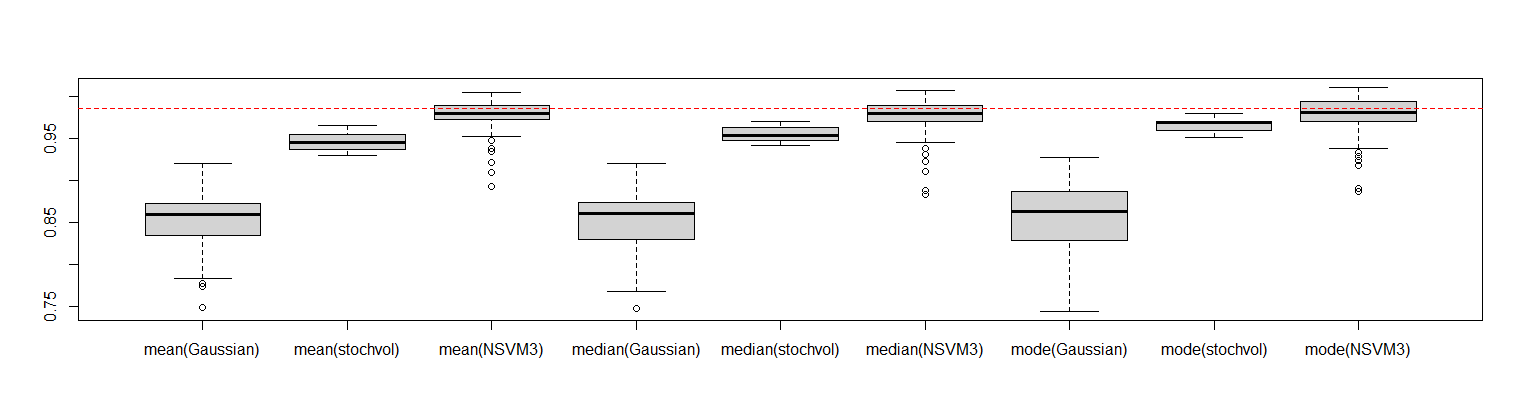}
    \caption{Delta estimation (Dependent Student's t error simulation)}
    \label{fig:100run_delta_t}
\end{figure}

\begin{figure}[H]
    \centering
    \includegraphics[width=\linewidth]{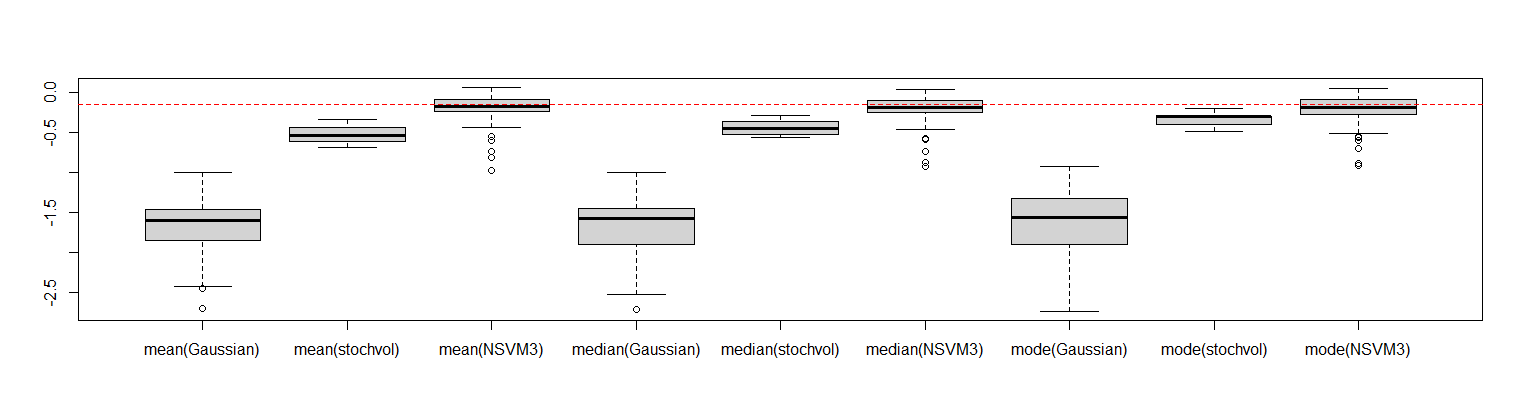}
    \caption{Alpha estimation (Dependent Student's t error simulation)}
    \label{fig:100run_alpha_t}
\end{figure}

\begin{figure}[H]
    \centering
    \includegraphics[width=\linewidth]{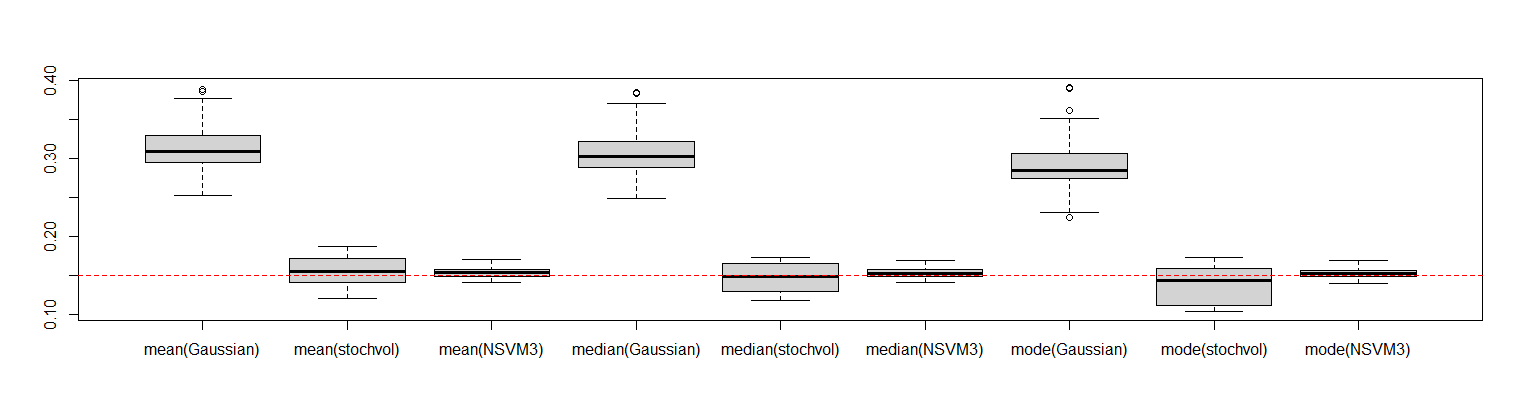}
    \caption{Sigma estimation (Dependent Student's t error simulation)}
    \label{fig:100run_sigma_t}
\end{figure}

The result of dependent Student's t simulation is similar to dependent Gaussian simulation; the NSVM-3 exhibits a smaller deviation from the red dashed line, representing the true parameter value. Again, we use the root mean square error(srMSE) to assess the deviation quantitatively. Table \ref{tab:para_est_compare_t} shows the srMSE of mean, median, and mode of the posterior of parameter $\delta$, $\alpha$, and $\sigma_\nu$ from 100 runs.

\begin{table}[H]
    \centering
    \resizebox{\textwidth}{!}{
    \begin{tabular}{|c|c|c|c|c|c|c|c|c|c|c|}
        \hline
        Model  & $\delta$ Mean & $\delta$ Median & $\delta$ Mode & $\alpha$ Mean & $\alpha$ Median & $\alpha$ Mode & $\sigma_\nu$ Mean & $\sigma_\nu$ Median & $\sigma_\nu$ Mode \\
        \hline
        Gaussian &  0.1378194  & 0.1366683 & 0.1347003 & 1.566643 & 1.554747 & 1.531064 & 0.1653401 & 0.1581516 & 0.1446725 \\
        stochvol  &  0.0409015 & 0.0323815 & 0.0209847 & 0.396245 & 0.313659 & 0.200964 & 0.0196587 & 0.0181448 & 0.0258145 \\
        NSVM-3  &  \textbf{0.01970314}  & \textbf{0.0236352} & \textbf{0.0240432} & \textbf{0.177015} & \textbf{0.180345} & \textbf{0.1912295} & \textbf{0.0074155} & \textbf{0.0071542} & \textbf{0.0067244} \\
        \hline
    \end{tabular}}
    \caption{Parameter estimation comparison, dependent Student's t error simulation (lower is better)}
    \label{tab:para_est_compare_t}
\end{table}

The quantitative results are consistent with the visual evidence presented in Figures \ref{fig:100run_delta_t}, \ref{fig:100run_alpha_t}, and \ref{fig:100run_sigma_t}, where the NSVM-3 model exhibits the lowest square root mean squared error (srMSE), indicating minimal bias in parameter estimation.

In the following section, we compare the volatility estimates generated by each algorithm with the true simulated volatility to evaluate the differences in their performance in volatility estimation.

\subsection{Volatility Estimation}

\begin{figure}[H]
    \centering
    \includegraphics[width=\linewidth]{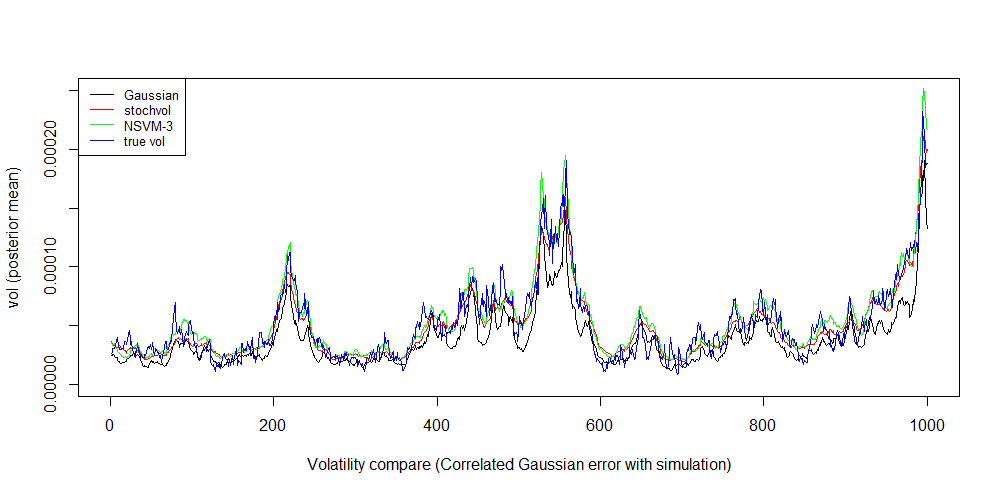}
    \caption{Volatility comparison (Dependent Gaussian error simulation)}
    \label{fig:vol_compare_Gaussian}
\end{figure}

To evaluate the model's effectiveness in estimating volatility, we create a dataset using the same parameters outlined earlier: $\alpha= -0.10$, $\delta = 0.985$, $\sigma = 0.15$. The algorithm was executed with identical data inputs for 100 repetitions. Each run utilized the MCMC algorithm for 10,000 iterations, with the initial 5,000 iterations designated as burn-in. Consequently, only the latter half of the samples, totaling 5,000 volatility values drawn from the posterior distribution, were retained. The mean of these posterior volatility samples was computed to obtain the estimated volatility. This procedure employed Gaussian error simulation, and the estimated volatility from the Gaussian model, as well as the \textsc{stochvol} and NSVM-3 models, is presented in Figure \ref{fig:vol_compare_Gaussian}.

\begin{figure}[H]
    \centering
    \includegraphics[width=\linewidth]{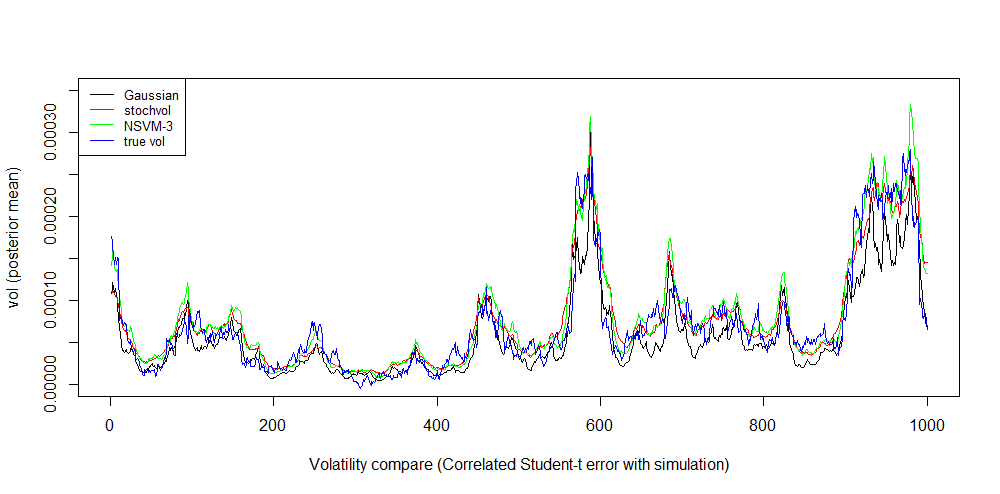}
    \caption{Volatility comparison (Dependent Student-t error simulation)}
    \label{fig:vol_compare_t}
\end{figure}

For Student-t error simulation, the comparison is shown in Figure \ref{fig:vol_compare_t}.

Additionally, we employ the following metrics to evaluate and compare the accuracy of volatility estimation. Let the true volatility values from the simulation be denoted by $h_1,...,h_N$ and the corresponding estimates by $\bar{h_1},...,\bar{h_N}$. In addition to considering the mean, we incorporate the median and mode of the final sampled volatilities for comparison. Specifically, we use three measures: the square root of the mean squared error (srMSE), $\sqrt{\frac{\Sigma (h_i - \bar{h_i})^2}{N}}$, the mean absolute error (MAE), $\frac{\Sigma |h_i - \bar{h_i}|}{N}$, and mean absolute percent error (MAPE), $\frac{1}{N} \Sigma 
 \frac{|h_i - \bar{h_i}|}{h_i}\times 100\%$. The table \ref{tab:vol_compare_error_gaussian_corr} presents the results of the Gaussian error simulation. Table \ref{tab:vol_compare_error_t_corr} displays the outcomes of the Student's t error simulation.

\begin{table}[H]
    \centering
    \resizebox{\textwidth}{!}{
    \begin{tabular}{|c|c|c|c|c|c|c|c|c|c|c|}
        \hline
        Model  & srMSE Mean & srMSE Median & srMSE Mode & MAE Mean & MAE Median & MAE Mode & MAPE Mean & MAPE Median & MAPE Mode \\
        \hline
        Gaussian &  0.008278  & 0.008287 & 0.008296 & 0.007433 & 0.007341 & 0.007347 & \textbf{0.09932} & \textbf{0.09908} & 0.09922 \\
        stochvol  &  0.008252  & 0.008260 & 0.008271 & 0.007313 & 0.007325 & 0.007327 & 0.09945 & 0.09927 & \textbf{0.09912} \\
        NSVM-3  &  \textbf{0.008250}  & \textbf{0.008259} & \textbf{0.008269} & \textbf{0.007213} & \textbf{0.007318} & \textbf{0.007314} & 0.09959 & 0.09939 & 0.09944 \\
        \hline
    \end{tabular}}
    \caption{Volatility estimation error comparison (dependent Gaussian error simulation)}
    \label{tab:vol_compare_error_gaussian_corr}
\end{table}

\begin{table}[H]
    \centering
    \resizebox{\textwidth}{!}{
    \begin{tabular}{|c|c|c|c|c|c|c|c|c|c|c|}
        \hline
        Model  & srMSE Mean & srMSE Median & srMSE Mode & MAE Mean & MAE Median & MAE Mode & MAPE Mean & MAPE Median & MAPE Mode \\
        \hline
        Gaussian &  0.006675  & 0.006682 & 0.006691 & 0.006019 & 0.006025 & 0.006031 & 0.09959 & 0.09968 & 0.09977 \\
        stochvol  &  \textbf{0.006648}  & 0.006656 & 0.006677 & \textbf{0.005997} & 0.006005 & 0.006011 & \textbf{0.09948} & \textbf{0.09957} & \textbf{0.09965} \\
        NSVM-3  &  0.006651  & \textbf{0.006651} & \textbf{0.006668} & 0.005998 & \textbf{0.006003} & \textbf{0.006009} & 0.09949 & 0.09959 & 0.09966 \\
        \hline
    \end{tabular}}
    \caption{Volatility estimation error comparison (dependent Student-t error simulation)}
    \label{tab:vol_compare_error_t_corr}
\end{table}

The tables \ref{tab:vol_compare_error_gaussian_corr}, \ref{tab:vol_compare_error_t_corr} indicate that NSVM-3 generally produces lower square root mean squared error (srMSE), mean absolute error (MAE), and mean absolute percentage error (MAPE). These reduced levels of bias offer assurance that NSVM-3 offers enhanced accuracy in estimating volatility.

\section{Empirical application}

This section presents the empirical findings obtained by applying our model to daily stock return data. We focus on the S$\&$P 500 (Standard and Poor’s 500), a widely recognized stock market index that tracks the performance of 500 of the largest publicly traded companies in the United States. As one of the most established indicators of the U.S. stock market and the broader economy, the S$\&$P 500 offers a suitable dataset for illustrating the robustness of our approach. Here, we utilize daily closing prices of it from February 1, 2021, through February 1, 2024; therefore, the number of trading days is $n=505$.

Our simulation study demonstrates that the proposed model yields both sampled volatility and parameter estimates. Specifically, we compare three models—the Gaussian model, \textsc{stochvol}, and NSVM-3, using the log returns of the closing price (\(\log \left( Price_{t}/Price_{t-1} \right)\)) as the input data. Each model is executed 100 times, employing a Markov Chain Monte Carlo (MCMC) procedure with 10,000 iterations, while discarding the first 5,000 iterations as burn-in. Thus, all subsequent inferences are based on the last 5,000 retained samples. We kept the prior setting the same in the previous section.

\begin{figure}[H]
    \centering
    \includegraphics[width=\linewidth]{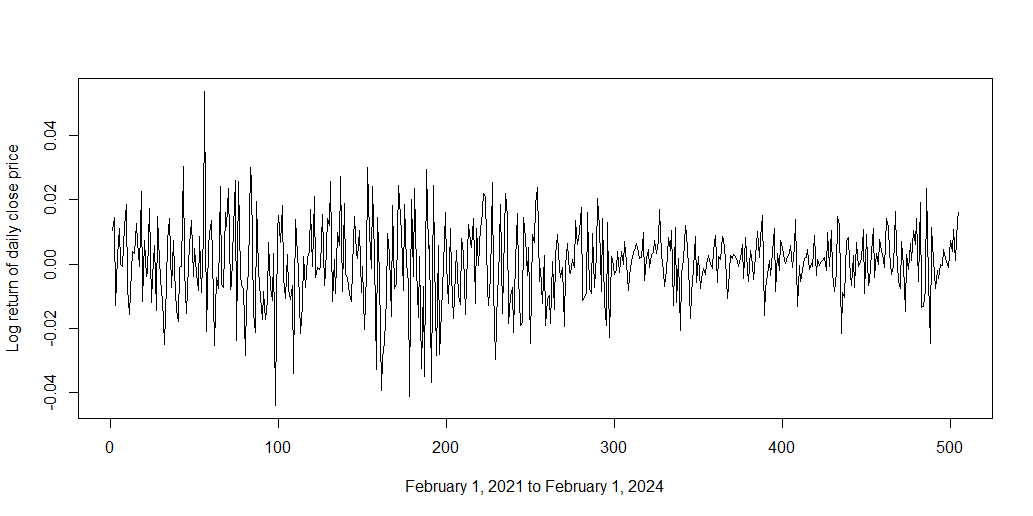}
    \caption{Daily log return of S$\&$P 500 series}
    \label{fig:return_sp}
\end{figure}

\begin{figure}[H]
    \centering
    \includegraphics[width=\linewidth]{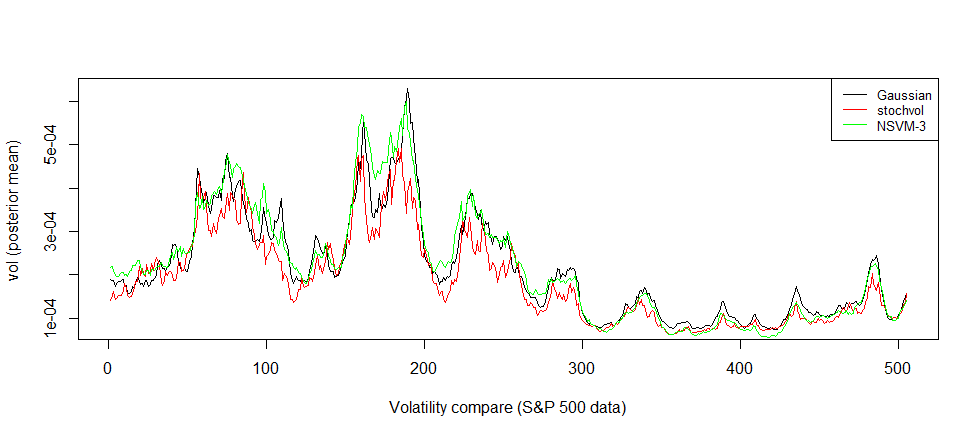}
    \caption{Comparison of volatility estimates of S$\&$P 500 series}
    \label{fig:vol_est_sp}
\end{figure}

Figure \ref{fig:return_sp} illustrates the logarithm of the return on the closing price for that period. In contrast, Figure \ref{fig:vol_est_sp} displays the average recorded volatility derived from the Gaussian model, \textsc{stochvol} method, and NSVM-3 model. Notably, the estimated volatility exhibits a marked increase at time points $t=60$ and $t=190$, responding to substantial fluctuations in return values, whether gains or losses. This behavior demonstrates that the estimated volatility effectively captures the magnitude of price movements, thus reflecting the intensity of these changes.

\begin{figure}[H]
    \centering
    \includegraphics[width=\linewidth]{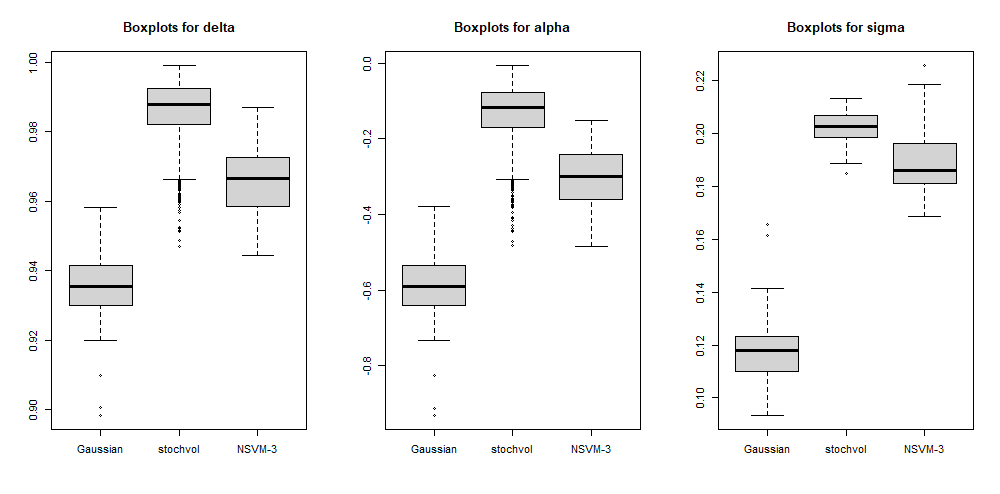}
    \caption{Box-plots of parameter estimates, S$\&$P 500 data}
    \label{fig:param_est_sp_corr}
\end{figure}

For parameter estimation, we record the means of the posterior distributions, $p(\alpha|h,\delta,\sigma_\nu^2)$, $p(\sigma_\nu^2|h,\alpha,\delta)$, and $p(\delta|h,\alpha,\sigma_\nu^2)$. Figure \ref{fig:param_est_sp_corr} displays the box plots of the means based on 100 repetitions of parameter estimates.

Identifying the true values of parameters in practical settings is considerably more challenging than in simulated environments, where such values are predefined and readily obtainable. This complexity stems from the multifaceted nature of real-world data, often influenced by various non-financial factors.

\section{Discussion and conclusion}

This paper introduces a semiparametric approach to modeling stochastic volatility (SV) that employs nonparametric estimation of all error distributions, thereby addressing the limitations imposed by Gaussian assumptions in conventional parametric models. It is one step further compared to the work by \cite{semiSVM}, which does not assume $u_t$ and $\nu_t$ are independent. This represents a significant relaxation of the assumptions regarding the innovation terms in the stochastic volatility model. By adopting less restrictive assumptions, the model allows for greater flexibility in capturing the underlying structure of the data, so data can ``speak'' for itself, thereby reducing both bias and variance and yielding more accurate volatility estimates.

The comprehensive simulation results underscore the advantages of our semiparametric method, which allows correlation between error terms, compared to the traditional Gaussian or bivariate Gaussian models. Utilizing the nonparametric numerical method and the Bayesian inference framework, the method is able to capture the non-Gaussian and possibly dependent features observed in financial time series data, ultimately enhancing both the flexibility and the accuracy of the SV model.

The empirical analysis conducted using S$\&$P 500 daily return data underscores the practical applicability and robustness of the proposed semiparametric framework, which reinforces the model’s capacity to reflect real-world financial dynamics. However, the primary contribution of this study lies in its rigorous validation of the proposed methodology through a series of controlled simulation experiments. These simulations provide a clear and systematic comparison between the semiparametric models and conventional approaches, particularly under conditions involving heavy-tailed and dependent error structures. The results consistently demonstrate that the semiparametric models yield superior performance in terms of bias, variance, and overall estimation accuracy. By offering both theoretical and empirical justification, this study advances the literature on stochastic volatility modeling and highlights the benefits of relaxing parametric assumptions in favor of more flexible, data-driven approaches.

A noticeable feature of this framework is its ability to flexibly capture non-Gaussian characteristics, such as skewness, heavy tails, and intricate dependence structures, that often arise in real-world financial time series. The method offers enhanced adaptability and delivers more accurate inferences about volatility and model parameters by sidestepping strict parametric assumptions. While the proposed method addresses key shortcomings of standard SV models, several potential research areas could be pursued for future work. The estimation of posterior likelihood is performed using the kernel density method, which, in certain cases, can be computationally intensive. Developing a more computationally efficient approach for estimating probability likelihood could significantly enhance the overall speed and efficiency of the procedure. Furthermore, this study utilized the Metropolis-Hastings algorithm as the primary Markov Chain Monte Carlo (MCMC) sampling method. However, investigating the impact of alternative sampling techniques that require fewer computing resources to improve the method's efficiency remains an important area for future research.

\newpage
\bibliography{Reference}

\newpage

\section{Appendix}

\subsection{Detailed algorithm}
The following are detailed algorithms that are mentioned in the main text.

\begin{algorithm}
    \begin{algorithmic}
        \Require $Input: (y,v_0,s_0,\delta_0,\sigma_\delta^2,\alpha_0,\sigma_\alpha^2,T,b)$
        \State $N \gets length(y)$
        \State $\delta \gets 1$
        \State $\alpha \gets 0$
        \State $\sigma_\nu^2 \gets 0.1$
        \State $h \gets rep(var(y),N)$
        \State $ln\_h \gets log(h)$
        \State $h\_sample \gets matrix(N,t)$
        \State $\delta\_sample \gets rep(0,t)$
        \State $\alpha\_sample \gets rep(0,t)$
        \State $\sigma_\nu^2\_sample \gets rep(0,t)$
        \For{$ite\ in\ 1\ to\ (T+b)$}
            \For{$i\ in\ 2\ to\ N-1$}
                \State $r \gets sample\_h(y[i],\alpha,\delta,\sigma_\nu^2,ln\_h[i-1],ln\_h[i+1])$
                \State $h[i] \gets r\$ h\_new$
                \State $ln\_h \gets log(h[i])$
            \EndFor
            \State $ln\_h[1] \gets rnorm(1,\alpha+\delta*ln\_h[2],\sqrt{\sigma_\nu^2})$
            \State $ln\_h[N] \gets rnorm(1,\alpha+\delta*ln\_h[N-1],\sqrt{\sigma_\nu^2})$
            \State $h[1] \gets exp(ln\_h[1])$
            \State $h[N] \gets exp(ln\_h[N])$
            \State $s_1 \gets sum(ln\_h)$
            \State $s_2 \gets sum(ln\_h^2)$
            \State $s_3 \gets sum(ln\_h[1:(N-1)]*ln\_h[2:N])$
            \State $s' \gets s_0+(N-1)\alpha^2+(1+\delta^2)s_2-\delta^2(ln\_h[N])^2-ln\_h[1]^2-2\alpha((1-\delta)s_1-ln\_h[1]+\delta ln\_h[N])-2\delta s_3$
            \State $\sigma_\nu^2 \gets sample\_\sigma_\nu^2(v_0,s',\alpha,\delta,h,lhleft,\sigma_\nu^2)$
            \State $\alpha \gets sample\_\alpha(\alpha_0,\sigma_\alpha^2,\delta,\sigma_\nu^2,h,lhleft,\alpha)$
            \State $\delta \gets sample\_\delta(\delta_0,\sigma_\delta^2,\alpha,\sigma_\nu^2,h,lhleft,\delta)$
            \If{$ite>b$}
                \State $\alpha\_sample[ite-b] \gets \alpha$
                \State $\delta\_sample[ite-b] \gets \delta$
                \State $\sigma_\nu^2\_sample[ite-b] \gets \sigma_\nu^2$
                \State $h\_sample[ite-b] \gets h$
            \EndIf
        \EndFor
        \State \Return $(h\_sample,\alpha\_sample,\delta\_sample,\sigma_\nu^2\_sample)$
    \end{algorithmic}
    \caption{Main}
    \label{Alg:sampler_main}
\end{algorithm}

\begin{algorithm}
    \begin{algorithmic}
        \Require $Input: (y,\alpha,\delta,\sigma_\nu^2,ln\_h_{i-1},ln\_h_{i+1})$
        \State $\mu \gets \frac{\delta (lhleft+lhright)+(1+\delta)\alpha}{1+\delta^2}$
        \State $\sigma^2 \gets \frac{\sigma_\nu^2}{1+\delta^2}$
        \State $\lambda \gets \frac{1-2exp(\sigma^2)}{1-exp(\sigma^2)}+0.5$
        \State $\phi \gets (\lambda-1)exp(\mu+\frac{\sigma^2}{2})+\frac{y^2}{2}$
        \State $qmode = \frac{\phi}{1+\lambda}$
        \State $ln\_c \gets log(1.1)+ln\_p(y,qmode,\alpha,\delta,\sigma_\nu^2)-log(dinvgamma(qmode,\lambda,\phi))$
        \State $h\_new \gets rinvgamma(1,\lambda,\phi)$
        \State $accept1 \gets \frac{exp(ln\_p(y,h\_new,qmode,\alpha,\delta,\sigma_\nu^2)-ln\_c)}{dinvgamma(h\_new,\lambda,\phi)}$
        \State $roll \gets runif(1)$
        \While{$roll>accept1$}
            \State $h\_new \gets rinvgamma(1,\lambda,\phi)$
            \State $accept1 \gets \frac{exp(ln\_p(y,h\_new,qmode,\alpha,\delta,\sigma_\nu^2)-ln\_c)}{dinvgamma(h\_new,\lambda,\phi)}$
            \State $roll \gets runif(1)$
        \EndWhile
        \If{$accept1 \geq 1$}
            \State $accept2 \gets exp(ln\_p(y,h\_new,qmode,\alpha,\delta,\sigma_\nu^2)-ln\_p(y,h,qmode,\alpha,\delta,\sigma_\nu^2)+log(dinvgamma(h,\lambda,\phi))-log(dinvgamma(h\_new,\lambda,\phi)))$
            \State $roll \gets runif(1)$  
            \If{$roll>accept2$}
                \State $h\_new \gets h$
            \EndIf
        \EndIf
        \State \Return $h\_new$
    \end{algorithmic}
    \caption{Sampler for $h$}
    \label{alg:sampler_h}
\end{algorithm}

\begin{algorithm}
    \begin{algorithmic}
        \Require $Input: (v_0,s',\alpha,\delta,h,ln\_h_{i-1},\sigma_{left}^2)$
        \State $q(\sigma_\nu^2) \sim IG(\frac{v_0+N-1}{2},\frac{s'}{2})$
        \State $p(\sigma_\nu^2) \gets p(\sigma_\nu^2|h,\alpha,\delta)$
        \State $c \gets 1.1*(\frac{p(\sigma_\nu^2)}{q(\sigma_\nu^2)})$, at $\sigma_\nu^2$ = mode of $q(\sigma_\nu^2)$,i.e. $\frac{s'/2}{\frac{v_0+N-1}{2}+1}$
        \State $ln\_c \gets log(1.1)+log(p(\sigma_\nu^2))-log(q(\sigma_\nu^2))$
        \State $\sigma_\nu^2\_new \gets rinvgamma(1,\frac{v_0+N-1}{2},\frac{s'}{2})$
        \State $accept1 \gets \frac{exp(p(\sigma_\nu^2)-ln\_c)}{dinvgamma(\sigma_\nu^2\_new,\frac{v_0+N-1}{2},\frac{s'}{2})}$
        \State $roll \gets runif(1)$
        \While{$roll>accept1$}
            \State $\sigma_\nu^2\_new \gets rinvgamma(1,\frac{v_0+N-1}{2},\frac{s'}{2})$
            \State $accept1 \gets \frac{exp(p(\sigma_\nu^2)-ln\_c)}{dinvgamma(\sigma_\nu^2\_new,\frac{v_0+N-1}{2},\frac{s'}{2})}$
            \State $roll \gets runif(1)$
        \EndWhile
        \If{$accept1 \geq 1$}
            \State $accept2 \gets \frac{p(\sigma_\nu^2\_new)/p(\sigma_\nu^2\_left)}{dinvgamma(\sigma_\nu^2\_new,\frac{v_0+N-1}{2},\frac{s'}{2})/dinvgamma(\sigma_\nu^2\_left,\frac{v_0+N-1}{2},\frac{s'}{2})}$
            \State $roll \gets runif(1)$  
            \If{$roll>accept2$}
                \State $\sigma_\nu^2\_new \gets \sigma_\nu^2\_left$
            \EndIf
        \EndIf
        \State \Return $\sigma_\nu^2\_new$
    \end{algorithmic}
    \caption{Sampler for $\sigma_\nu^2$}
    \label{alg:sampler_sigma}
\end{algorithm}

\begin{algorithm}
    \begin{algorithmic}
        \Require $Input: (\delta_0,\sigma_\delta^2,\alpha,\sigma_\nu^2,h,ln\_h_{i-1},\delta_{left})$
        \State $qfunc\_mean\_\delta \gets \frac{\sigma_\nu^2\delta_0+\sigma_\delta^2(s_3-\alpha(S_1-\ln h_N))}{\sigma_\nu^2+\sigma_\delta^2(s_2-(\ln h_N)^2)}$
        \State $qfunc\_var\_\delta \gets \frac{\sigma_\nu^2\sigma_\delta^2}{\sigma_\nu^2+\sigma_\delta^2(s_2-(\ln h_N)^2)}$
        \State $q(\delta) \sim N\left(qfunc\_mean\_\delta,qfunc\_var\_\delta\right)$
        \State $p(\delta) \gets p(\delta|h,\alpha,\sigma_\nu^2)$
        \State $c \gets 1.1*(\frac{p(\delta)}{q(\delta)})$, at $\delta$ = mode of $q(\delta)$
        \State $ln\_c \gets log(1.1)+log(p(\delta))-log(q(\delta))$
        \State $\delta\_new \gets rnorm(1,qfunc\_mean\_\delta,qfunc\_var\_\delta)$
        \State $accept1 \gets \frac{exp(p(\delta)-ln\_c)}{dnorm(\delta\_new,qfunc\_mean\_\delta,qfunc\_var\_\delta)}$
        \State $roll \gets runif(1)$
        \While{$roll>accept1$}
            \State $\delta\_new \gets rnorm(1,qfunc\_mean\_\delta,qfunc\_var\_\delta)$
            \State $accept1 \gets \frac{exp(p(\delta)-ln\_c)}{dnorm(\delta\_new,qfunc\_mean\_\delta,qfunc\_var\_\delta)}$
            \State $roll \gets runif(1)$
        \EndWhile
        \If{$accept1 \geq 1$}
            \State $accept2 \gets \frac{p(\delta\_new)/p(\delta\_left)}{dnorm(\delta\_new,qfunc\_mean\_\delta,qfunc\_var\_\delta)/dnorm(\delta\_left,qfunc\_mean\_\delta,qfunc\_var\_\delta)}$
            \State $roll \gets runif(1)$  
            \If{$roll>accept2$}
                \State $\delta\_new \gets \delta\_left$
            \EndIf
        \EndIf
        \State \Return $\delta\_new$
    \end{algorithmic}
    \caption{Sampler for $\delta$}
    \label{alg:sampler_delta}
\end{algorithm}

\begin{algorithm}
    \begin{algorithmic}
        \Require $Input: (\alpha_0,\sigma_\alpha^2,\delta,\sigma_\nu^2,h,ln\_h_{i-1},\alpha_{left})$
        \State $qfunc\_mean\_\alpha \gets \frac{\sigma_\alpha^2((1-\delta)s_1-\ln h_1+\delta\ln h_N)+\sigma_\nu^2\alpha_0}{\sigma_\nu^2+(N-1)\sigma_\alpha^2}$
        \State $qfunc\_var\_\alpha \gets \frac{\sigma_\nu^2\sigma_\alpha^2}{\sigma_\nu^2+(N-1)\sigma_\alpha^2}$
        \State $q(\alpha) \sim N\left(qfunc\_mean\_\alpha,qfunc\_var\_\alpha\right)$
        \State $p(\alpha) \gets p(\alpha|h,\delta,\sigma_\nu^2)$
        \State $c \gets 1.1*(\frac{p(\alpha)}{q(\alpha)})$, at $\alpha$ = mode of $q(\alpha)$
        \State $ln\_c \gets log(1.1)+log(p(\alpha))-log(q(\alpha))$
        \State $\alpha\_new \gets rnorm(1,qfunc\_mean\_\alpha,qfunc\_var\_\alpha)$
        \State $accept1 \gets \frac{exp(p(\alpha)-ln\_c)}{dnorm(\alpha\_new,qfunc\_mean\_\alpha,qfunc\_var\_\alpha)}$
        \State $roll \gets runif(1)$
        \While{$roll>accept1$}
            \State $\alpha\_new \gets rnorm(1,qfunc\_mean\_\alpha,qfunc\_var\_\alpha)$
            \State $accept1 \gets \frac{exp(p(\alpha)-ln\_c)}{dnorm(\alpha\_new,qfunc\_mean\_\alpha,qfunc\_var\_\alpha)}$
            \State $roll \gets runif(1)$
        \EndWhile
        \If{$accept1 \geq 1$}
            \State $accept2 \gets \frac{p(\alpha\_new)/p(\alpha\_left)}{dnorm(\alpha\_new,qfunc\_mean\_\alpha,qfunc\_var\_\alpha)/dnorm(\alpha\_left,qfunc\_mean\_\alpha,qfunc\_var\_\alpha)}$
            \State $roll \gets runif(1)$  
            \If{$roll>accept2$}
                \State $\alpha\_new \gets \alpha\_left$
            \EndIf
        \EndIf
        \State \Return $\alpha\_new$
    \end{algorithmic}
    \caption{Sampler for $\alpha$}
    \label{alg:sampler_alpha}
\end{algorithm}

\end{document}